\documentclass[a4paper,11pt]{article}
\usepackage{pos}

\renewcommand{\logo}{\relax}

\title{TeV bayesian study of the extragalactic background light}

\author*{Lucas Gréaux}
\author{Jonathan Biteau}

\affiliation{Université Paris-Saclay, CNRS/IN2P3, IJCLab, \\
    91405 Orsay, France}

\emailAdd{lucas.greaux@ijclab.in2p3.fr}
\emailAdd{jonathan.biteau@ijclab.in2p3.fr}

\newcommand{\Phiobs}{\Phi_{\rm obs}}
\newcommand{\Phiint}{\Phi_{\rm int}}
\newcommand{\diff}{\rm d}
\newcommand{\comment}[1]{}

\def\P(#1){\Phelper#1|\relax\Pchoice(#1)}
\def\Phelper#1|#2\relax{\ifx\relax#2\relax\def\Pchoice{\Pone}\else\def\Pchoice{\Ptwo}\fi}
\def\Pone(#1){\Pr\left( #1 \right)}
\def\Ptwo(#1|#2){\Pr\left( #1 \,\middle|\, #2 \right)}
\def\Pr{\mathbf{Pr}}

\def\D{\mathcal{D}}

\abstract{
    The extragalactic background light (EBL) is the aggregate of all optical and infrared emissions from thermal processes since the cosmic dark ages. While the integrated light of galaxies is expected to be the main contribution to the EBL, recent measurements beyond Pluto’s orbit from the New Horizon probe show a 4$\sigma$ excess in the optical band. This tension can be studied within observational gamma-ray cosmology, by reconstructing EBL-induced absorption features in the gamma-ray spectra of extragalactic sources at very-high energies (VHE, $E>100\,$GeV). Gamma-ray studies of the EBL remain limited by the size of the spectral corpora and by the uncertainties on the shape of the spectra emitted at the sources. We developed a new analysis method that aims to tackle these limitations. Unlike existing studies, we employ a fully Bayesian framework, which allows us to remove arbitrary criteria for selecting intrinsic spectral models. Such an approach further enables marginalization over systematics of instrumental origin, such as the uncertainty on the energy scale of current-generation VHE observatories. In this contribution, we apply our method to the most extensive catalog of extragalactic VHE spectra to date, STeVECat. We present preliminary constraints on the energy density of the EBL at redshift $z=0$, obtained with 259 archival VHE spectra from 56 extragalactic sources with known redshift.
}

\FullConference{%
  7th Heidelberg International Symposium on High-Energy Gamma-Ray Astronomy (Gamma2022)\\
  4-8 July 2022\\
  Barcelona, Spain\\}

\begin{document}
\maketitle

\section{Introduction}

\comment{
Trucs à dire:
\begin{itemize}
    \item Change number in abstract
    \item Put number of sources from STeVECat
\end{itemize}
}

The cosmic voids are filled with photons that form diffuse backgrounds.
The most intense background of astrophysical origin is the extragalactic background light (EBL, see \cite{Gamma-ray_cosmo}), which is defined here as all optical and infrared emissions since the cosmic dark ages.
The main contribution to the EBL is expected to be the integrated light of galaxies (IGL), which can be measured thanks to deep galaxies surveys \cite{K21}.
Low-surface brightness emissions from galaxies or from their surroundings are less known than their brighter counterparts, and IGL studies could tend to underestimate the intensity of the EBL.
The EBL can also be measured directly, by observing the brightness of dark patches in the sky \cite{M19}.
These measurements can be contaminated by foreground emissions like sunlight reflection on interplanetary dust: direct estimates tend to overestimate the intensity of the EBL.

Beyond Pluto's orbit, the interplanetary dust glow is expected to be negligible. 
Recent direct measurement from the New Horizons probe beyond Pluto's orbit show a 4$\sigma$ excess in the optical band with respect to the IGL measurements \cite{L22}.
This tension can be studied within observational gamma-ray cosmology \cite{Gamma-ray_cosmo}.
Above an energy threshold which depends on redshift, very-high energy (VHE, $E > 100\,$GeV) photons can interact with photons from the EBL to produce electron-positron pairs.
Observed spectra of extragalactic sources show EBL-induced absorption features at TeV energies that can be reconstructed to measure the intensity of the EBL, with a typical resolution of 20\% as of today \cite{Gamma-ray_cosmo}.

Gamma-ray studies of the EBL face two main challenges.
The VHE spectra emitted by extragalactic sources are unknown and the uncertainties on their shape can affect EBL reconstruction \cite{Biasuzzi_2019}. 
Furthermore, EBL-induced absorption depends both on the energy of the emitted gamma rays and on the source's redshift.
VHE studies require large spectral corpora to grasp these dependencies. 
To tackle these limitations, we developed a new analysis method for gamma-ray studies of the EBL.
Unlike existing studies, we employ a fully Bayesian framework which takes into account the uncertainties on the spectral shape of the sources as well as systematic uncertainties of instrumental origin.
In this contribution, we apply our method to the most extensive catalog of extragalactic VHE spectra to date, the Spectral TeV Extragalactic Catalog (STeVECat, see \cite{STeVECat}).
We present preliminary constraints on the energy density of the EBL at redshift $z=0$, obtained with 259 archival VHE spectra from 56 extragalactic sources with $z > 0.01$.

\section{Analysis framework} \label{sec:pipeline}

\subsection{Spectral reconstruction}

The interaction between VHE gamma-rays and the EBL is characterized by the EBL optical depth, $\tau(E, z)$, which depends on the energy $E$ and redshift $z$ at which the VHE photons were emitted (see \cite{Gamma-ray_cosmo}).
We parameterize here the EBL optical depth through a scaling factor, $\alpha$, with respect to a reference model $\tau_{\rm ref}$.
The observed spectrum, $\Phiobs$, of an extragalactic source located at redshift $z$ can be written as $\Phiobs(E, z) = \Phiint(E)\times e^{-\alpha\cdot\tau_{\rm ref}(E, z)}$, where $\Phiint$ is the intrinsic spectrum we would observe without absorption on the EBL.
For this parameterization, $\alpha=0$ means that no absorption is seen, while $\alpha=1$ means that the absorption features are fully compatible with the reference model $\tau_{\rm ref}$.

The intrinsic spectra of the sources are unknown.
They are expected to follow power-laws, with or without intrinsic curvature, and the choice of intrinsic spectral model can affect the EBL reconstruction \cite{Biasuzzi_2019}.
Most studies use frequentist inference and proceed by iteration over a corpus of spectral models to select the best sets of models describing the ensemble of observations.
We chose to use a Bayesian framework, which allows us to consider arbitrarily complex spectral models and treat their parameters as nuisance parameters.
By reconstructing the complete probability distribution of the parameters instead of finding the most probable value, we are able to compensate the over-fitting effects induced by the increase in complexity of the spectral model.

We model the intrinsic spectra of the sources as log-parabolae with an exponential cutoff.
Following \cite{Nigro_2019}, we take into account the uncertainties on the energy scale of the observing telescopes by adding an additional parameter $\epsilon = \frac{E'}{E} - 1$, where $E$ is the true energy of an event, while $E'$ is observed energy, reconstructed by the instrument.

\begin{figure}[t]
    \centering
    \includegraphics[width=0.8\linewidth]{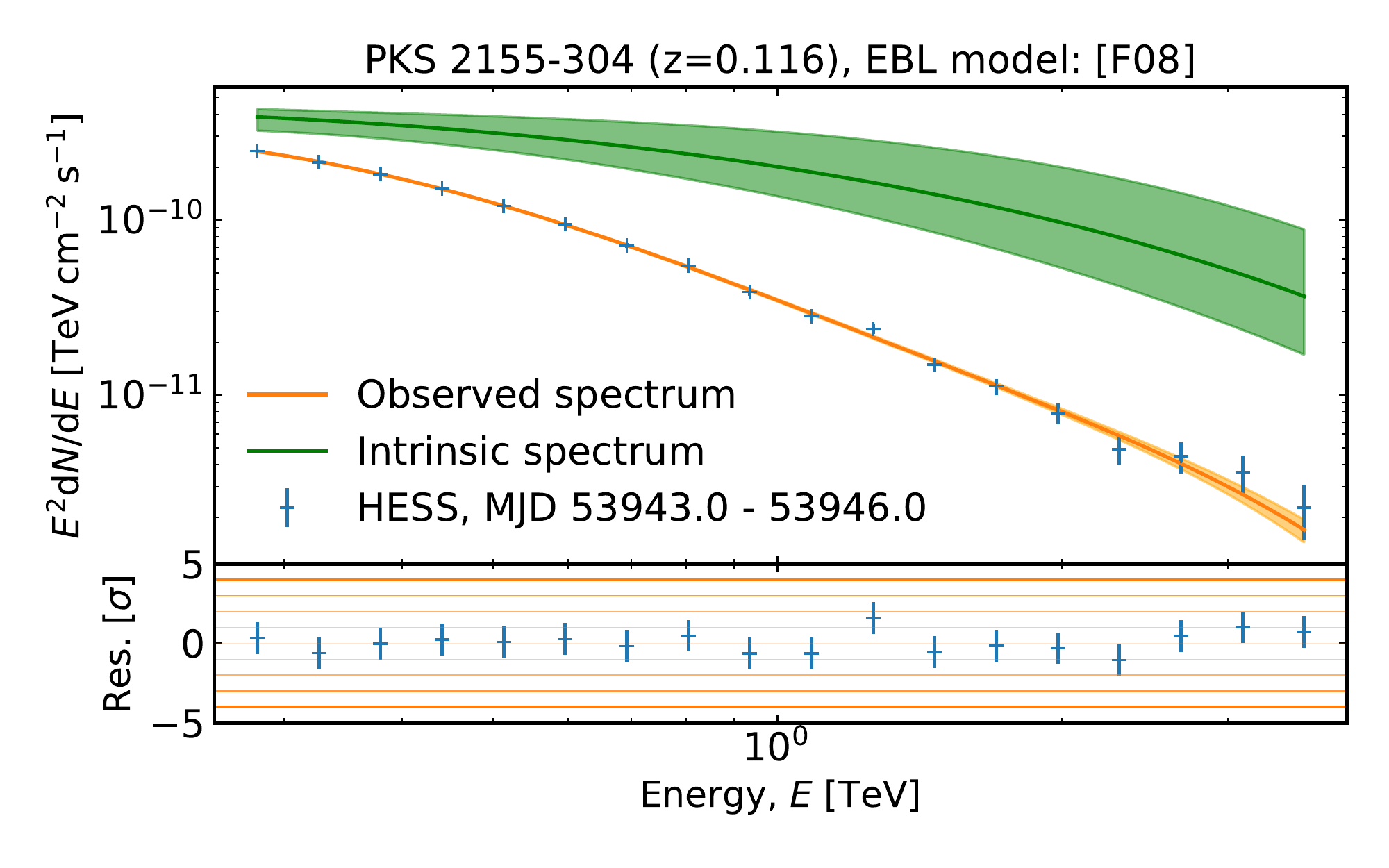}
    \caption{Reconstructed spectrum for the source PKS 2155-304 (\emph{top}) and associated residuals (\emph{bottom}), using the observations from H.E.S.S. between MJD 53943 and MJD 53946 \cite{2013PhRvD..88j2003A}. The H.E.S.S. data points are represented in blue. The orange line corresponds to the reconstruction of the observed spectrum which accounts for the EBL absorption from F08 (see text); and the green line corresponds to the reconstructed intrinsic spectrum. The bands correspond to the associated $1\sigma$ uncertainties.}
    \label{fig:spectrum}
\end{figure}

Considering a spectrum of index $i$ with observed spectral points $D_i$ and spectral parameters $\Theta_i$, we use Markov chain Monte Carlo to compute the posterior distribution $\P(\alpha, \Theta_i| D_i)$ from the likelihood $\P(D_i | \alpha, \Theta_i)$ and priors on the parameters $\alpha$ and $\Theta_i$, $\P(\alpha)$ and $\P(\Theta_i)$:\footnote{
We use non-informative prior distributions for the log-parabola parameters to minimize the \textit{a priori} knowledge on the intrinsic spectra.
We use a flat prior for the EBL normalization parameter $\alpha$, and we use a gaussian prior for the energy bias parameter $\epsilon$ following \cite{Nigro_2019}.}
\begin{eqnarray} \label{eq:bayes_formula}
    \P(\alpha, \Theta_i| D_i) & = & \frac{\P(D_i | \alpha, \Theta_i)\P(\alpha)\P(\Theta_i)}{\P(D_i)} {\rm .}
\end{eqnarray}
One example of a reconstructed spectrum is presented in Fig.~\ref{fig:spectrum}, which corresponds to a flare from the blazar PKS 2155-304 observed by H.E.S.S. in July, 2006 \cite{2013PhRvD..88j2003A}.

\subsection{EBL reconstruction}

From each $\P(\alpha, \Theta_i| D_i)$ defined in Eq.~\ref{eq:bayes_formula}, we can extract a probability distribution for the normalization factor $\alpha$ by marginalizing on the parameters $\Theta_i$, that is by computing $\P(\alpha | D_i) = \int \diff \Theta_i~\P(\alpha, \Theta_i| D_i)$. 
It is then possible to compute the probability distribution of $\alpha$ given all observations $\D = \{D_i\}_i$:
\begin{eqnarray} \label{eq:proba_product}
    \P(\alpha|\D) & = & \frac{\P(\D|\alpha)\P(\alpha)}{\P(\D)} = \P(\alpha) \prod_i \frac{\P(D_i|\alpha)}{\P(D_i)}\nonumber \\
        & = & \P(\alpha) \prod_i \frac{\P(\alpha|D_i)}{\P(\alpha)} {\rm .}
\end{eqnarray}

\begin{figure}[t]
    \centering
    \includegraphics[width=0.8\linewidth]{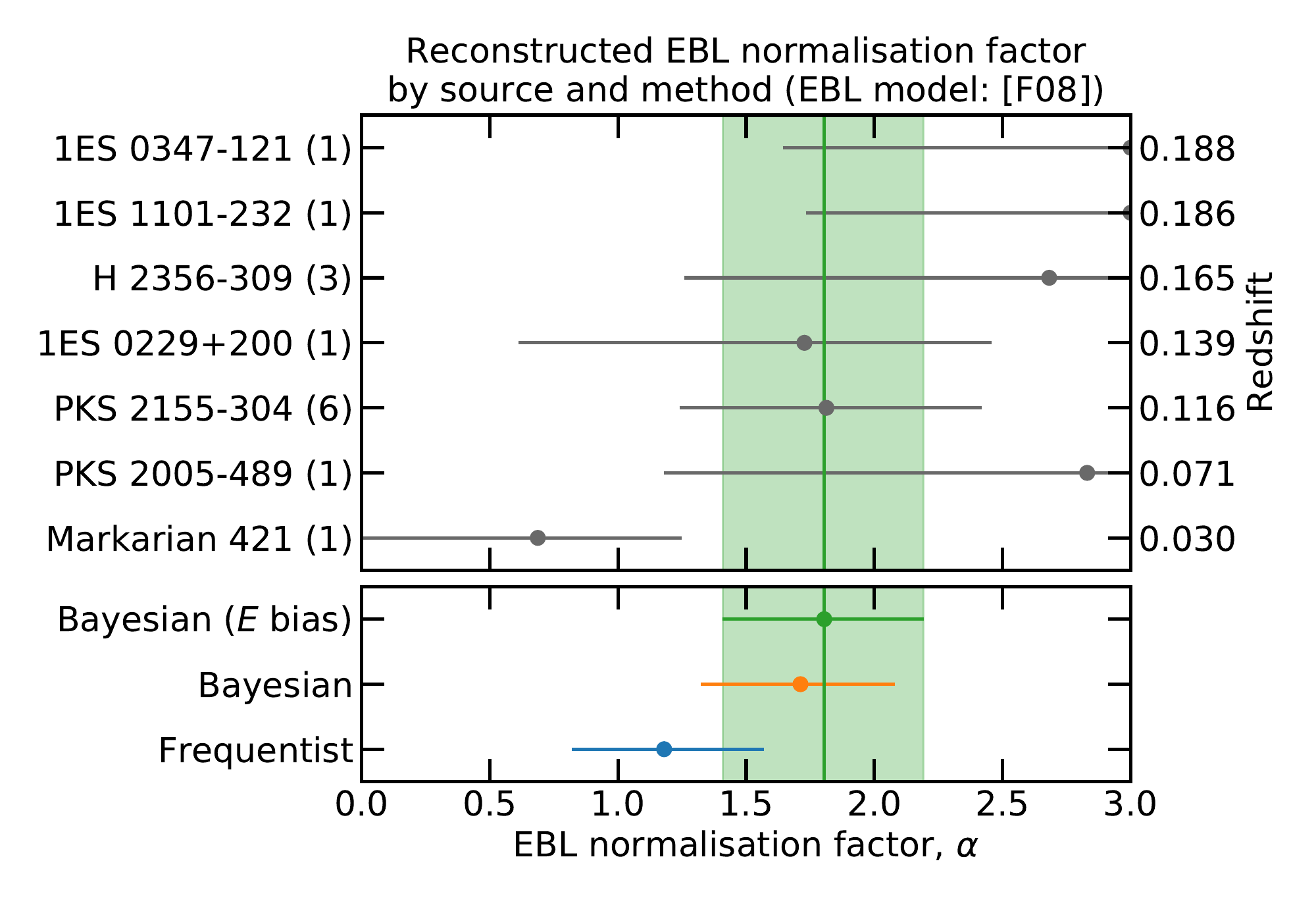}
    \caption{Reconstructed EBL normalization factor for each source (\emph{top}) using the Bayesian reconstruction with energy-bias, and global normalization factor using different reconstruction methods (\emph{bottom}) using the EBL model from F08 (see text). For each source, the associated number of spectra is written in parenthesis.}
    \label{fig:sources}
\end{figure}

We present in Fig.~\ref{fig:sources} the results of this procedure on a set of 14 spectra observed by H.E.S.S. with high significance, following \cite{Biteau_2013}.
We use the obsolete EBL model from \cite{F08} (here F08) for the sake of comparison with previous publications.
We display the results from this Bayesian pipeline with and without the energy scale parameter $\epsilon$, as well as an implementation of a frequentist analysis following \cite{Biasuzzi_2019}.
The source-by-source reconstruction in the top panel corresponds to the Bayesian reconstruction using a free parameter $\epsilon$.

The normalization factors $\alpha$ reconstructed with the three methods are compatible with the value reported in \cite{Biteau_2013}.
The energy scale bias factor we reconstruct is around 8\%, which is consistent with the values reported in \cite{Nigro_2019}.

\section{Using the STeVECat corpus}

The Spectral TeV Extragalactic Catalog (STeVECat, see \cite{STeVECat}) is the most comprehensive collection of VHE spectra to date, with more than 400 spectra published in journal articles.
To constrain the intensity from the optical part of the EBL, we select 259 spectra from STeVECat from sources with known redshift $z > 0.01$.
We apply the procedure developed in Sec.~\ref{sec:pipeline}, using the energy scale bias as free parameter $\epsilon$.

\begin{figure}[t]
    \centering
    \includegraphics[width=0.8\linewidth]{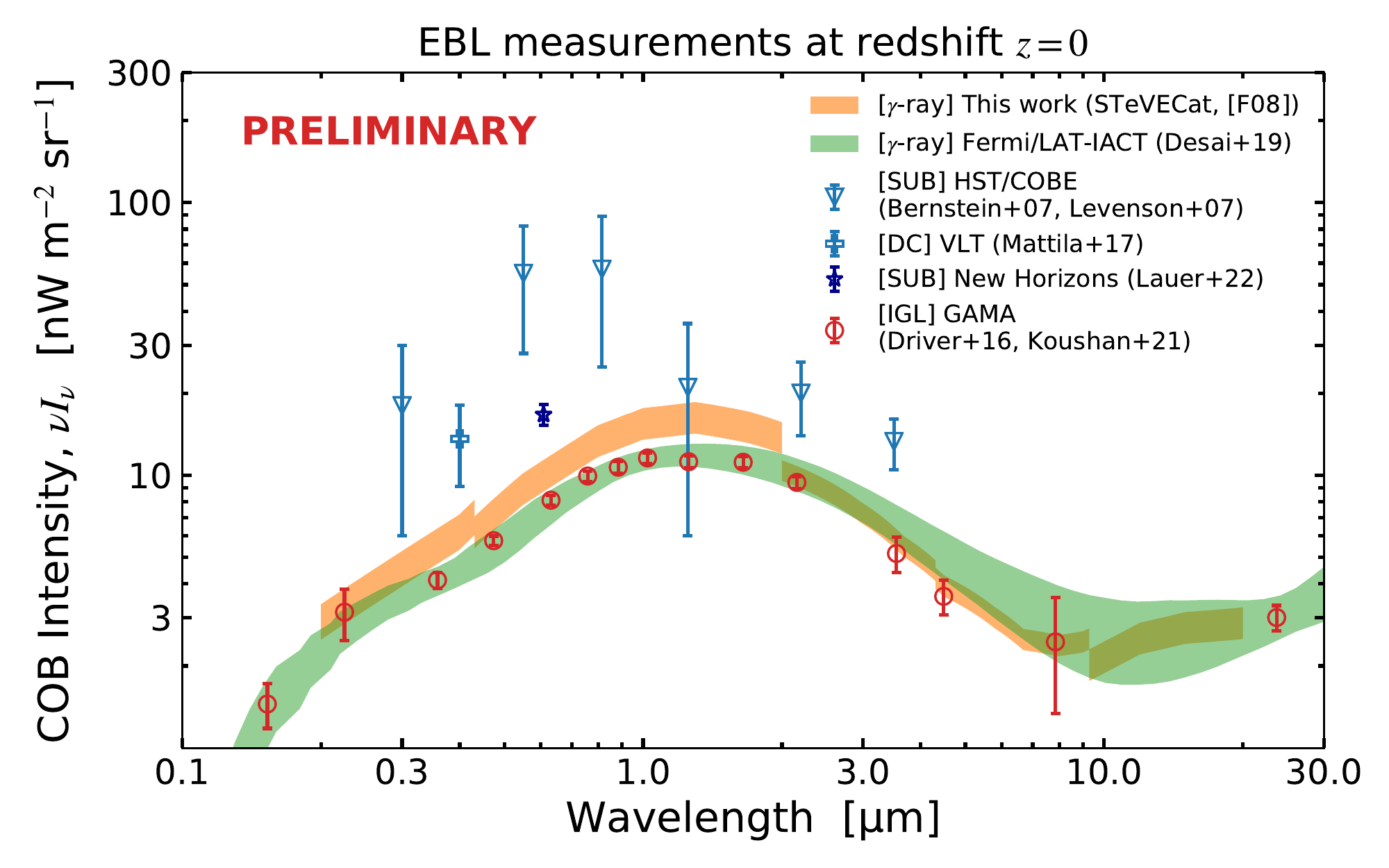}
    \caption{EBL intensity spectrum at redshift $z=0$. The orange band corresponds to the reconstruction from this work, using the STeVECat dataset and the EBL model from F08 (see text). The green band corresponds to the EBL measurement from \cite{D19}. The blue points corresponds to direct measurements (see \cite{M19, L22}) and the red points corresponds to IGL measurements (see \cite{K21}).}
    \label{fig:ebl}
\end{figure}

We present in Fig.~\ref{fig:ebl} our EBL measurement using the STeVECat corpus and considering six wavelength bins from 0.2 to 20~$\mu$m.
For each bin, we reconstructed a normalization factor following Eq.~\ref{eq:proba_product} considering only the spectra expected to be affected by the EBL at these wavelengths, as in \cite{Biteau_2013}.
The resulting measurement corresponds to the orange band, which is compatible with previous gamma-ray studies.

The typical resolution on gamma-ray measurements of the EBL is around 20\% \cite{Gamma-ray_cosmo}.
We are able to improve this resolution to around 15\% while only using VHE observations.
This preliminary EBL level is compatible with IGL measurements, but the six considered energy bins are not independent.
To solve the optical controversy, we are developing a wavelength-resolved reconstruction.

\section{Conclusion}

Studies of the EBL using gamma rays are moving from an era of discovery to an era of precision. 
Such studies are limited by the lack of knowledge on the intrinsic spectra of the gamma-ray sources and by the size of the available spectral samples.
We developed a new framework to reconstruct the EBL intensity using the Bayesian formalism, which allows us to use a general intrinsic spectral model in order to remove arbitrary model selection criteria.
This approach further enables marginalization over systematics of instrumental origin, such as the uncertainty on the energy scale of current-generation VHE observatories.

We applied this new framework to a corpus of 259 spectra from the most extensive VHE spectral catalog, STeVECat.
We obtained a measurement of the EBL intensity in the optical band compatible with current IGL measurements, bringing down uncertainties of gamma-ray cosmology measurements to 15\%.
To further improve this reconstruction, we are currently working on the inclusion of contemporaneous \textit{Fermi}-LAT measurements in the GeV band to better constrain the spectra emitted by the sources. 
We are also developing a model-independent EBL parametrization, which could allow us to constrain the origin of the optical excess evidenced by New Horizons.

\bibliographystyle{JHEP}
\bibliography{arxiv}

\end{document}